\documentclass{article}
\usepackage{geometry}
\usepackage{amsmath}
\usepackage{amssymb}
\usepackage{graphicx}
\usepackage{color}

\DeclareMathOperator{\Tr}{Tr}

\newcommand{\half}{\textstyle\frac{1}{2}}
\newcommand{\dd}{\mathrm{d}} 
\newcommand{\free}{\alpha}

\newcommand{\bfXi}{\boldsymbol{\Xi}}
\newcommand{\bfxi}{\boldsymbol{\xi}}
\newcommand{\mbf}[1]{{\boldsymbol{#1}}}
\newcommand{\tbf}[1]{{\textbf{#1}}}

\newcommand{\bK}{\mathbf{K}}
\newcommand{\bX}{\mathbf{\times}}
\newcommand{\bdot}{\mathbf{\cdot}}
\newcommand{\bom}{\boldsymbol{\omega}}

\begin{document}

\begin{center}
\begin{Large}
{\bf Nonholonomic Noetherian symmetries and integrals \\ of the Routh sphere 
     {and Chaplygin ball}}
\end{Large}

\medskip
\begin{large}
{\bf Miguel D. Bustamante and Peter Lynch}\\
\emph{\small School of Mathematics and Statistics, University College Dublin, Belfield, Dublin 4, Ireland.}
\end{large}
\end{center}

\bigskip

\bigskip
\centerline{\bf{\sc Abstract}}

\medskip
\begin{narrower}
\begin{small}
The dynamics of a spherical body with a non-uniform mass distribution
rolling on a plane were discussed by Sergey Chaplygin, whose 150$^\mathrm{th}$
anniversary we celebrate this year. The {Chaplygin top}
is a non-integrable system, with a colourful range of interesting
motions. A special case of this system was studied by Edward Routh,
who showed that it is integrable.

The Routh sphere has centre of mass offset from the geometric centre,
but it has an axis of symmetry through both these points,
and equal moments of inertia about all axes orthogonal to the symmetry axis.
There are three constants of motion: the total energy and two quantities
involving the angular momenta.

It is straightforward to demonstrate that these quantities,
known as the Jellett and Routh constants, are integrals of the motion.
However, their physical significance has not been fully understood.
In this paper, we show how the integrals of the Routh sphere arise
from Emmy Noether's invariance identity.  \textcolor{black}{We derive expressions for
the infinitesimal symmetry transformations associated with these constants.
We find the finite version of these symmetries and  provide their geometrical interpretation.} 

{As a further demonstration of the power and utility of this method,
we find the Noether symmetries and corresponding Noether integrals for a system
introduced recently: the Chaplygin ball on a rotating turntable, confirming that the
known integrals are directly obtained from Noether's theorem. }

\end{small}
\end{narrower}


\section{Introduction}

This paper is a contribution to the celebration of the 150th anniversary
of the birth of Sergey Alexeyevich Chaplygin (1869--1942), the renowned
Russian physicist, mathematician, and mechanical engineer.  Amongst many
other topics, Chaplygin studied the dynamics of a sphere rolling on a
plane. For this {\emph{Chaplygin top}},
the mass distribution is eccentric, the three moments of inertia are
distinct, and the geometric centre does not, in general, lie on any of
the principal axes.

A special case of this system was studied by Edward Routh \cite{Routh05}.
The Routh sphere is a spherical body with a non-uniform distribution
of mass, free to roll without slipping on a plane surface.  Its centre of
mass is offset from the geometric centre, but it has an axis of symmetry
through both these points, and equal moments of inertia about all axes
orthogonal to the symmetry axis.  This distinguishes it from the more
general case studied by Chaplygin \cite{Chaplygin03}.

Routh \cite{Routh05} showed that the Routh sphere has two constants of motion
in addition to the energy, and is an integrable system.  The integrals
or constants of motion, known as Jellett's constant and Routh's
constant, have been treated in many studies. 
We mention, in particular, the important contributions 
\cite{BoMa15, Cushman98, Byungsoo11, Koslov02}.
A simple proof that Jellett's and Routh's constants are
integrals of the motion is given in Gray and Nickel \cite{GrayNickel00}.
However, as remarked by these authors, ``The
precise physical significance of the Routh constant remains elusive \dots\
[and] it might be useful to try to find a direct connection between this
constant of the motion and the underlying symmetries of the system''
\cite[p.~826]{GrayNickel00}.  This explicit connection is established
in the present work.

Emmy Noether discovered a fundamental connection between symmetries or
invariances of dynamical systems and conserved quantities or integrals of
the motion.  For a historical review, see \cite{KS11}.  In her seminal
paper \cite{Noether18}, Noether derived an identity, valid whenever the
action of the system has an invariance.  In the case of extremal flow,
in which the Euler-Lagrange or d'Alembert-Lagrange equations apply, this
leads to a Noetherian conservation law.  This is true both for systems
with holonomic constraints and for systems with nonholonomic constraints
that are linear in the velocities.  We will show in this paper how the
integrals of the Routh sphere arise from Noether's invariance identity,
and will derive expressions for the symmetry transformations associated
with these constants.

\textcolor{black}{
As a further demonstration of the power and utility of Noether's theorem,
we examine in \S\ref{sec:ChapBall} the problem of the Chaplygin ball on a
rotating turntable, recently studied in \cite{BBM18}. Using a systematic
approach, we deduce the four known integrals and their associated symmetries
directly from Noether's invariance identity.}


\section{The invariance identity}

Associated with invariance of the action functional under transformations
of the dependent and independent variables there is an identity,
the \emph{invariance identity.} \textcolor{black}{We restrict ourselves,
at the expense of generality but for simplicity of presentation, to the case
when the transformation does not depend on the velocities.}
Then the invariance identity may be expanded in
powers of the velocity variables \textcolor{black}{$\dot q^\mu, \,\,\mu=1, \ldots, N$,
where $N$ is the number of degrees of freedom}, to yield a set of
differential equations. If these can be solved, they provide the
generators of a coordinate transformation that can be used to
construct a constant of the motion.

For a dynamical system with a Lagrangian function, 
let us define the action functional
$$
S = \int_{t_1}^{t_2} L\left(q(t),\frac{\dd q(t)}{\dd t},t\right)\, \dd t \,.
$$
We consider a continuous transformation of the independent and dependent variables
\textcolor{black}{$$
t \to T(q,t;\free), \qquad q^\mu \to Q^\mu(q,t;\free)\,,
$$
where $\free \in \mathbb{R}$ is a free parameter.  The case $\free = 0$ corresponds
to the identity transformation, with $T(q,t;0) = t$ and $Q^\mu(q,t;0) = q^\mu$.
We form the action $S^\prime$ using the new variables but
the same functional form of the Lagrangian $L$:
$$
S^\prime = \int_{T_1}^{T_2} L\left(Q(T),\frac{\dd Q(T)}{\dd T},T\right)\,\dd T\,,
$$
where $Q^\mu(T)$ (with slight abuse of notation) stands for the new variable as a function of the new time. 
We consider the case where the action is invariant under the transformation: $S^\prime = S$.}
For an infinitesimal perturbation, we write 
$$\begin{array}{ccccccc}
q^\mu(t) &\longrightarrow& Q^\mu(T) &=& q^\mu(t) &+& \epsilon\, \xi^\mu(q,t) \,,\\
t &\longrightarrow& T &=& t &+& \epsilon\, \tau(q,t) \,.
\end{array}$$
The coefficients of $\epsilon$ are called the generators of the transformation. \textcolor{black}{They form the components of a vector field $\left(\xi^\mu(q,t) , \tau(q,t)\right)$, called an infinitesimal Noether symmetry.}
We expand the integrand of $S^\prime$ and express it as an integral
with respect to $t$. Then the following \relax{invariance identity} results:
\begin{equation}
\frac{\partial L}{\partial q^\mu}\xi^\mu + p_\mu \dot\xi^\mu +
\frac{\partial L}{\partial t}\tau - H \dot\tau = 0
\label{eq:InvId1} 
\end{equation}
where $p_\mu={\partial L}/{\partial\dot q^\mu}$ is the conjugate momentum,
the Hamiltonian is $H = p_\mu \dot q^\mu - L$, and 
the Einstein summation convention is employed. 
This identity was first derived by Emmy Noether \cite{Noether18}.
Eq.~(\ref{eq:InvId1})
can be written in a completely equivalent but more illuminating form:
\begin{equation}
\frac{\dd}{\dd t}\biggl[p_\mu \xi^\mu - H\tau \biggr]
= (\xi^\mu-\dot q^\mu \tau) 
   \left[ \frac{\dd}{\dd t}\frac{\partial L}{\partial\dot q^\mu} 
        - \frac{\partial L}{\partial q^\mu} \right]\,.
\label{eq:InvId2}
\end{equation}


\subsection*{Extremal or on-shell motion}

\textcolor{black}{The term in square brackets on the right hand side of 
Eq.~(\ref{eq:InvId2}) is the Euler-Lagrange operator acting on the Lagrangian:
\[E_{\mu}[L] \equiv
\frac{\dd}{\dd t}\frac{\partial L}{\partial\dot q^\mu} 
- \frac{\partial L}{\partial q^\mu} \,.
\]}
For a holonomic system, this expression vanishes, so the following conservation law holds:
\begin{equation}
 \frac{\dd}{\dd t}\biggl[p_\mu \xi^\mu - H\tau \biggr] = 0\,.
\label{eq:Noether1}
\end{equation}

For a general nonholonomic system, little can be said. However, if the $M$ constraints
are linear in the velocities, so that
 \textcolor{black}{$$
\gamma^\kappa \equiv A^\kappa_\mu(q,t) \dot q^\mu + B^\kappa(q,t) = 0\,, \quad \kappa = 1, \ldots, M,
$$}
then the d'Alembert-Lagrange equations may be written in the form
$$
\left[ \frac{\dd}{\dd t}\frac{\partial L}{\partial \dot q^\mu} - \frac{\partial L}{\partial q^\mu} \right]
= \lambda_\kappa \frac{\partial \gamma^\kappa}{\partial \dot q^\mu} = \lambda_\kappa A^\kappa_\mu  \,.
$$
The right hand side of Eq.~(\ref{eq:InvId2}) then becomes
\textcolor{black}{$$
(\xi^\mu-\dot q^\mu \tau)  \left[ \frac{\dd}{\dd t}\frac{\partial L}{\partial \dot q^\mu} - \frac{\partial L}{\partial q^\mu} \right]
= (\xi^\mu-\dot q^\mu \tau) \lambda_\kappa A^\kappa_\mu  = 
\lambda_\kappa (A^\kappa_\mu \xi^\mu + B^\kappa \tau)  \,.
$$}
If we assume that the \textcolor{black}{infinitesimal Noether symmetry} respects the constraints, namely if
\begin{equation}
A^\kappa_\mu \xi^\mu + B^\kappa \tau = 0,\qquad \kappa = 1, \ldots, M \,, 
\label{eq:symm_constraints}
\end{equation}
then this expression vanishes.
As a consequence, the right hand side of Eq.~(\ref{eq:InvId2}) vanishes for on-shell flow.

We conclude that, for both holonomic systems and systems subject to nonholonomic constraints
that are linear in the velocities, even with inhomogeneous terms, Eq.~(\ref{eq:InvId2})
reduces to the conservation law, Eq.~(\ref{eq:Noether1}) \cite{Bahar87}.


\section{Routh sphere}

\begin{figure}[h]
\begin{center}
\includegraphics[width=0.75\linewidth]{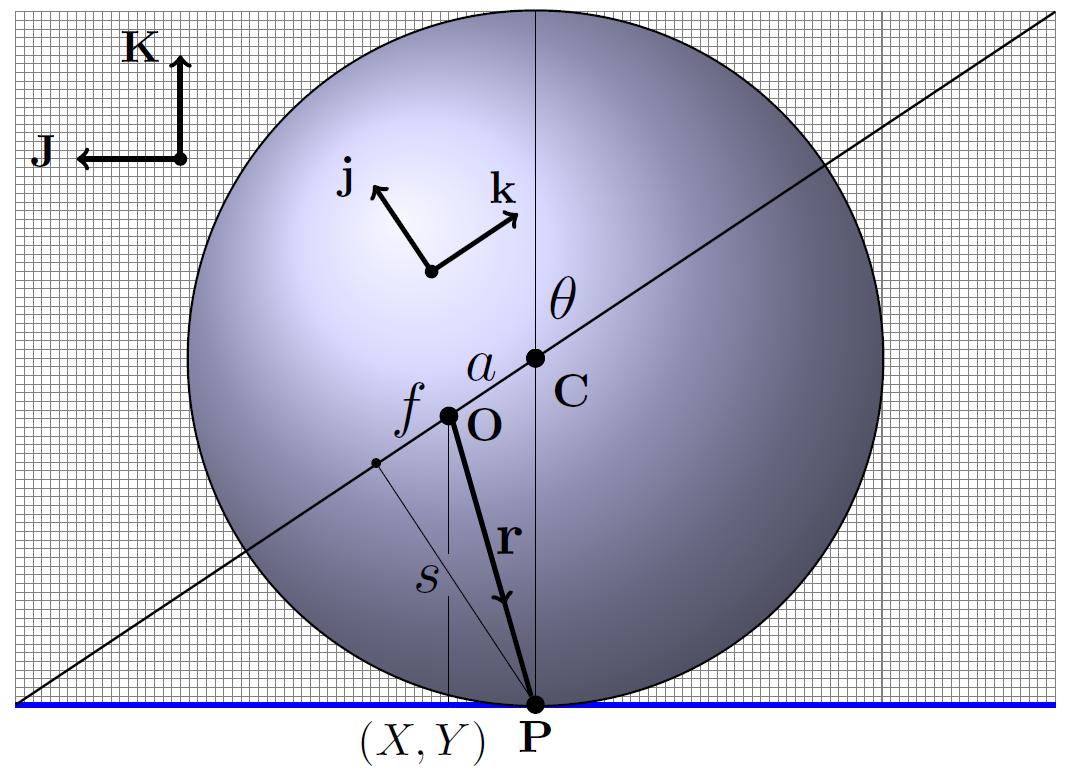}
\caption{Geometry and primary coordinates for the Routh sphere.
Geometric centre $\mathbf{C}$, mass centre $\mathbf{O}$ and point of contact $\mathbf{P}$.
In this configuration, $\mathbf{I}$ and $\mathbf{i}$ point into the page and
$\phi = -\pi/2$.}
\label{fig:RS}
\end{center}
\end{figure}

The dynamics of the Routh sphere are discussed in many texts on classical mechanics.
The original study is \cite{Routh05}. In this paper we follow the notation of 
\cite{LyBu09} and \cite{LyBu13}. There are  six degrees of freedom: the configuration of the body is
given by $(X,Y,Z)$, the coordinates of the centre of mass, and the three Euler angles
$(\theta,\phi,\psi)$.
The unit orthogonal triad in the space frame is $\{ \mathbf{I}, \mathbf{J}, \mathbf{K} \}$
and the unit orthogonal triad in the intermediate frame is $\{ \mathbf{i}, \mathbf{j},
\mathbf{k} \}$ with $\mathbf{i}$ horizontal and $\mathbf{k}$ fixed along the axis of the
body (see Fig.~\ref{fig:RS}). 

\textcolor{black}{The holonomic constraint that the geometric centre must remain at unit distance above the underlying plane is used to eliminate the variable $Z$, leading to an effective system with $N=5$ degrees of freedom.} Assuming unit mass and unit radius, the Lagrangian of the Routh sphere is
\smallskip
\[
L = \half\bigl[(I_1+a^2s^2)\dot\theta^2 + (I_1 s^2+I_3 c^2 )\dot\phi^2 + (2I_3 c)\dot\phi\dot\psi +
(I_3)\dot\psi^2 + \dot X^2 + \dot Y^2 \bigr]  - ga(1-c)
\]
\smallskip\noindent
where $s=\sin\theta$, $c=\cos\theta$ and other notation is conventional.
We note that $L$ is independent of both $\phi$ and $\psi$.
We assume that $I_1 = I_2 \ne I_3$.

\textcolor{black}{There are $M=2$ nonholonomic constraints, which are linear and homogeneous in the velocities,
corresponding to rolling motion without slipping:}
\begin{eqnarray}
\dot X &=& \phantom{-}h s_\phi \dot\theta - a s c_\phi \dot\phi - s c_\phi \dot\psi 
\label{eq:Xdot} \\
\dot Y &=&          - h c_\phi \dot\theta - a s s_\phi \dot\phi - s s_\phi \dot\psi
\label{eq:Ydot}
\end{eqnarray}
where $c_\phi=\cos\phi$, $s_\phi=\sin\phi$ and $h=1-ac$ is the height of the centre of mass.
We write these constraints in the form $\gamma^\kappa \equiv A^\kappa_\mu \dot q^\mu = 0$ where
$\dot q^\mu = \left(\dot\theta,\dot\phi,\dot\psi,\dot X, \dot Y\right)$ and
$$
A^\kappa_\mu = 
\left[
\begin{matrix}
- h s_\phi & asc_\phi & sc_\phi & 1 & 0 \\
  h c_\phi & ass_\phi & ss_\phi & 0 & 1
\end{matrix}
\right]\,.
$$
For reference, we note that
$$
\dot X^2 + \dot Y^2 = h^2\dot\theta^2 + s^2(a\dot\phi+\dot\psi)^2 \,.
$$
However, we cannot use this to eliminate $\dot X$ and $\dot Y$ from the Lagrangian
as the constraints are nonholonomic \cite{Flannery05}.


The conjugate momenta are defined in terms of the Lagrangian:
$p_\mu=\partial L/\partial \dot q^\mu$. For the Routh sphere they are 
\begin{eqnarray*}
p_\theta &=& (I_1+a^2s^2)\dot\theta \\
p_\phi   &=& (I_1 s^2+I_3 c^2 )\dot\phi + (I_3 c)\dot\psi \\
p_\psi   &=& (I_3 c )\dot\phi + (I_3)\dot\psi  \,.
\end{eqnarray*}
We also have $p_X = \dot X$ and $p_Y = \dot Y$. 
Since the determinant of the coefficients (the Hessian)
is $(I_1 + a^2s^2) I_1 I_3 s^2$, we can solve for the velocities:
\begin{eqnarray*}
\dot\theta &=& p_\theta / (I_1+a^2s^2) \\
\dot\phi   &=& (p_\phi -c p_\psi)/I_1 s^2 \\
\dot\psi   &=& (-c/I_1 s^2) p_\phi + ((I_1 s^2+I_3 c^2)/I_1 I_3 s^2) p_\psi 
\end{eqnarray*}
and, of course, $\dot X = p_X$ and $\dot Y = p_Y $.

\subsection*{Invariance}

We note that $\phi$, $\psi$, $X$ and $Y$ are all ignorable coordinates. 
Thus, $L$ is invariant with respect to infinitesimal variations of 
these coordinates.  For free-slip boundary conditions, where there are 
no constraints linking the momenta, there are four conserved quantities
\[
\bigl\{ p_\phi, p_\psi, p_X, p_Y \bigr\}
\]
corresponding to these four coordinates.

Since the Lagrangian does not depend explicitly on $t$, invariance under
a transformation of the form $t^\prime = t + \epsilon\tau$ \textcolor{black}{with $\tau$ constant} leads,
in the usual way, to conservation of the energy.
We therefore assume a transformation of the space coordinates,
\begin{eqnarray*}
{\phi}^\prime &=& \phi + \epsilon\, \xi^\phi(\theta) \\
{\psi}^\prime &=& \psi + \epsilon\, \xi^\psi(\theta) 
\end{eqnarray*}
where the generators are functions of $\theta$, so that
\[
\dot\xi^\phi = \frac{\dd\xi^\phi}{\dd\theta} \dot\theta
\qquad \mbox{and} \qquad 
\dot\xi^\psi = \frac{\dd\xi^\psi}{\dd\theta} \dot\theta \,.
\]
The constraints also require variations of $X$ and $Y$ of the form
\begin{eqnarray*}
{X}^\prime &=& X + \epsilon\, \xi^X(\theta,\phi) \\
{Y}^\prime &=& Y + \epsilon\, \xi^Y(\theta,\phi) 
\end{eqnarray*}
so that $\xi^X$ and $\xi^Y$ depend on $\phi$ as well as $\theta$.
Explicitly, the constraints imply
\begin{equation}
\xi^X = -s c_\phi (a \xi^\phi + \xi^\psi ) 
\qquad \mbox{and} \qquad 
\xi^Y = -s s_\phi (a \xi^\phi + \xi^\psi )  \,.
\label{eq:zetaXY}
\end{equation}
We note that $c_\phi\xi^X+s_\phi\xi^Y = -s(a\xi^\phi+\xi^\psi)$, independent of
$\phi$.
The time derivatives are
\begin{eqnarray*}
\dot\xi^X &=& 
\bigl[-c \,c_\phi(a\xi^\phi+\xi^\psi) -s c_\phi(a\xi^\phi_{,\theta}+\xi^\psi_{,\theta})\bigr] \dot\theta
+ \bigl[ s s_\phi(a\xi^\phi+\xi^\psi)\bigr]\dot\phi \\
\dot\xi^Y &=& 
\bigl[-c\, s_\phi(a\xi^\phi+\xi^\psi) -s s_\phi(a\xi^\phi_{,\theta}+\xi^\psi_{,\theta})\bigr] \dot\theta
- \bigl[ s c_\phi(a\xi^\phi+\xi^\psi)\bigr]\dot\phi \,.
\end{eqnarray*}
Again, we note that $c_\phi\dot\xi^X+s_\phi\dot\xi^Y$ is independent of $\phi$.
The invariance identity, Eq.~(\ref{eq:InvId1}), now becomes
\[
p_\phi \dot\xi^\phi + p_\psi \dot\xi^\psi + p_X \dot\xi^X + p_Y \dot\xi^Y = 0 \,.
\]
Substituting the above values we get, for the unconstrained variables,
\[
p_\phi \dot\xi^\phi + p_\psi \dot\xi^\psi =
[(I_1 s^2+I_3 c^2 )\xi^\phi_{,\theta} + (I_3 c )\xi^\psi_{,\theta} ]\dot\theta\dot\phi +
[(I_3 c )\xi^\phi_{,\theta} + (I_3 )\xi^\psi_{,\theta} ]\dot\theta\dot\psi
\]
and, for the constrained variables,
\[
p_X\dot\xi^X + p_Y\dot\xi^Y = 
\bigl[s(a\xi^\phi+\xi^\psi)+as^2(a\xi^\phi_{,\theta}+\xi^\psi_{,\theta})\bigr]\dot\theta\dot\phi +
\bigl[sc(a\xi^\phi+\xi^\psi)+s^2(a\xi^\phi_{,\theta}+\xi^\psi_{,\theta})\bigr]\dot\theta\dot\psi \,.
\]
Note that this expression is independent of $\phi$.
Adding these two expressions and setting the coefficients of $\dot\theta\dot\phi$ and
$\dot\theta\dot\psi$ separately to zero gives two ode's for $\xi^\phi$ and $\xi^\psi$:
\begin{eqnarray}
(I_1 s^2+I_3 c^2 + a^2 s^2 )\frac{\dd \xi^\phi}{\dd\theta} +
(I_3 c + a s^2 )\frac{\dd \xi^\psi}{\dd\theta} +
s(a\xi^\phi+\xi^\psi) &=& 0 \,,
\label{eq:sym1} \\
(I_3 c + a s^2 )\frac{\dd \xi^\phi}{\dd\theta} +
(I_3 + s^2 )\frac{\dd \xi^\psi}{\dd\theta} +
sc (a\xi^\phi+\xi^\psi) &=& 0\,.
\label{eq:sym2}
\end{eqnarray}
These are the symmetry equations for the Routh Sphere. 
We can write them
\begin{equation}
\mathsf{F}\,\frac{\dd{\bfxi}}{\dd\theta} = \mathsf{G}\,{\bfxi} 
\label{eq:gensys}
\end{equation}
where ${\bfxi} = (\xi^\phi,\xi^\psi)^{\mathrm T}$ and
the coefficient matrices are
\begin{equation*}
\mathsf{F} =
\begin{bmatrix}
I_1 s^2+I_3 c^2 + a^2 s^2   &   I_3 c + a s^2 \\
I_3 c + a s^2               &   I_3 + s^2
\end{bmatrix}
\qquad\mbox{and}\qquad
\mathsf{G} =
- \begin{bmatrix}
 a s    &  s   \\
 a s c  &  s c
\end{bmatrix} \,.
\end{equation*}
The determinant of the matrix $\mathsf{F}$ is $I_1 s^2/\rho^2$,
where
$$
\rho = \frac{1}{\sqrt{s^2+I_3+(I_3/I_1)f^2}} \,.
$$
So $\mathsf{F}$ is invertible and the symmetry equations may be written as
${\dd\bfxi}/{\dd\theta} = \mathsf{H}\,\bfxi$, where $\mathsf{H} = \mathsf{F}^{-1}\mathsf{G}$.
Explicitly,
\begin{equation}
\frac{\dd}{\dd\theta} 
\begin{pmatrix}
 \xi^\phi  \\
 \xi^\psi 
\end{pmatrix}
=
\left(-\frac{\rho^2 s}{I_1}\right)
\begin{bmatrix}
a(I_3 +h)             &   (I_3 + h)          \\
a(I_1 c - I_3 c -ha)  &  (I_1 c - I_3 c -ha)
\end{bmatrix}
\begin{pmatrix}
 \xi^\phi  \\
 \xi^\psi 
\end{pmatrix}\,.
\label{eq:Heqns}
\end{equation}

\subsection*{Solution of the symmetry equations}

One solution of Eqs.~(\ref{eq:sym1}) and (\ref{eq:sym2}) is immediately obvious by
inspection: take both $\xi^\phi$ and $\xi^\psi$ constant,
with $\xi^\phi = 1$ and $\xi^\psi = - a$.  Then $(a\xi^\phi+\xi^\psi)=0$ so,
by virtue of Eq.~(\ref{eq:zetaXY}), both $\xi^X$ and $\xi^Y$ vanish.
The Noetherian constant associated with this transformation is
\begin{equation}
C_J = p_\mu\xi^\mu = p_\phi - a p_\psi \,,
\label{eq:CJellet}
\end{equation}
which is Jellett's constant.

Once a solution of Eqs.~(\ref{eq:Heqns}) is known, another one can be found.  Suppose there are two linearly independent solutions
$(\xi^\phi_1,\xi^\psi_1)^{\rm T}$ and $(\xi^\phi_2,\xi^\psi_2)^{\rm T}$. The Wronskian is
defined to be the determinant
\[
W(\theta) = 
\left| \begin{matrix}
\xi^\phi_1   &  \xi^\phi_2  \\
\xi^\psi_1   &  \xi^\psi_2 
\end{matrix} \right|
=
\xi^\phi_1\xi^\psi_2 - \xi^\phi_2\xi^\psi_1  \,.
\]
It is easily shown that
\[
\frac{\dd W}{\dd \theta} = \Tr(\mathsf{H})\, W,
\]
where $\Tr(\mathsf{H}) = \mathsf{H}_{11} + \mathsf{H}_{22}$. This has a solution $W(\theta) = C \exp[\int \Tr(\mathsf{H})\,\dd \theta ]$. 
The explicit form of $\mathsf{H}$ is implied from 
Eq.~(\ref{eq:Heqns}) so that
$\Tr(\mathsf{H}) =  (-{\rho^2 s}/{I_1})[I_1 c - I_3(c-a)]$. This can be integrated to yield
$W(\theta)= C\rho$, with $C$ a constant depending on the normalisation choice for the linearly independent solutions. Then using the definition of $W$ we find that 
\[
\xi^\phi_2(\theta) = 
\xi^\phi_1(\theta) \int^\theta \frac{\mathsf{H}_{12}(\theta)}{\xi^\phi_1(\theta)^2} W(\theta)\,\dd\theta\,.
\]
In the present case, $\xi^\phi_1(\theta) = 1$, $\mathsf{H}_{12}(\theta) = (-\rho^2 s / I_1)(I_3+h)$ and we make the convenient choice $W(\theta)= I_1 \rho$.  We find, by direct integration, the solution 
$\xi^\phi_2(\theta) = (c-a)\rho$ and thence, since $W = a\xi^\phi_2+\xi^\psi_2$, we get
\[
\left( \begin{matrix}
\xi^\phi_2  \\ \xi^\psi_2
\end{matrix} \right)
=
\left( \begin{matrix}
f\rho  \\ (I_1-af)\rho 
\end{matrix} \right)\,,
\]
where we write $f = c - a$.  Eq.~(\ref{eq:zetaXY}) gives $\xi^X$ and $\xi^Y$.
Then the Noetherian constant is
\begin{equation}
C_R = p_\mu\xi^\mu = \left[\frac{I_1}{I_3}\right]\frac{p_\psi }{\rho}\,,
\label{eq:CRouth}
\end{equation}
which is Routh's constant.

We can now write the general solution of 
Eq.~(\ref{eq:Heqns}) as
\[
\left( \begin{matrix}
\xi^\phi  \\ \xi^\psi 
\end{matrix} \right)
=
A_1\left( \begin{matrix}
\xi^\phi_1  \\ \xi^\psi_1 
\end{matrix} \right)
+
A_2\left( \begin{matrix}
\xi^\phi_2  \\ \xi^\psi_2 
\end{matrix} \right)
=
\left( \begin{matrix}
A_1 + A_2 f\rho \\
-a A_1 + A_2(I_1-af)\rho
\end{matrix} \right)\,.
\]


\section{Recovering the symmetry from a known constant}

Suppose we know that $C=p_\mu\xi^\mu$ is a constant of the motion. Then
\begin{equation}
\frac{\partial p_\mu}{\partial p_\nu}\xi^\mu 
= \left[
  \frac{\partial p_\phi}{\partial p_\nu}\xi^\phi
+ \frac{\partial p_\psi}{\partial p_\nu}\xi^\psi
+ \frac{\partial p_X}{\partial p_\nu}\xi^X
+ \frac{\partial p_Y}{\partial p_\nu}\xi^Y \right]
= \frac{\partial C}{\partial p_\nu} 
\label{eq:dCdp}
\end{equation}
provides a system of equations for the generators $\xi^\mu$.
For unconstrained  motion the momenta are independent and it follows that
${\partial p_\mu}/{\partial p_\nu}=\delta_\mu^\nu$, so that
\[
\xi^\nu = \frac{\partial C}{\partial p_\nu} \,.
\]
For constrained motion, the generators are interconnected and a linear system of equations
must be solved.

We can write the constraints 
Eqs.~(\ref{eq:Xdot})--(\ref{eq:Ydot}) in terms of momenta:
\begin{eqnarray*}
p_X &=& \phantom{-}s_\phi\left(\frac{h}{I_1+a^2s^2}\right)p_\theta +
c_\phi\left[\left(\frac{f}{I_1s}\right)p_\phi-\left(\frac{fc}{I_1s}+\frac{s}{I_3}\right)p_\psi\right]\,,
\\
p_Y &=& -c_\phi\left(\frac{h}{I_1+a^2s^2}\right)p_\theta +
s_\phi\left[\left(\frac{f}{I_1s}\right)p_\phi-\left(\frac{fc}{I_1s}+\frac{s}{I_3}\right)p_\psi\right]\,.
\end{eqnarray*}
We also recall that the generators satisfy the constraints 
Eq.~(\ref{eq:zetaXY}):
\begin{equation*}
\xi^X = -s c_\phi (a \xi^\phi + \xi^\psi ) 
\qquad \mbox{and} \qquad 
\xi^Y = -s s_\phi (a \xi^\phi + \xi^\psi )  \,.
\end{equation*}
These expressions allow us to eliminate the momenta $p_X$ and $p_Y$ and the generators $\xi^X$ and $\xi^Y$
from 
Eq.~(\ref{eq:dCdp}) and obtain expressions relating $\xi^\phi$ and $\xi^\psi$:
\begin{eqnarray}
\xi^\phi - \left(\frac{f}{I_1}\right)(a\xi^\phi+\xi^\psi) 
&=& \frac{\partial C}{\partial p_\phi}
\label{eq:dCdp1} \\
\xi^\psi + \left(\frac{fc}{I_1}+\frac{s^2}{I_3}\right)(a\xi^\phi+\xi^\psi) 
&=& \frac{\partial C}{\partial p_\psi} \,.
\label{eq:dCdp2} 
\end{eqnarray}

Let us apply 
Eqs.~(\ref{eq:dCdp1})--(\ref{eq:dCdp2}) to the Jellett and Routh constants.
For the Jellett constant, $C_J=(p_\phi-ap_\psi)$, we have 
$({\partial C_J}/{\partial p_\phi},{\partial C_J}/{\partial p_\psi})=(1,-a)$ and
the solution is immediately obvious by inspection:
\begin{equation}
\bfXi_J \equiv
\left( \begin{matrix} \xi^\theta \\ \xi^\phi \\ \xi^\psi \\ \xi^X \\ \xi^Y \end{matrix}
\right) =
\left( \begin{matrix} 0  \\  1  \\  -a  \\  0  \\  0 \end{matrix} \right) \,.
\label{eq:Jgens}
\end{equation}
The coordinates $X$ and $Y$ of the centre of mass do not vary.
An interpretation of this vector, will be given in \S\ref{sec:interp} below.

For the Routh constant, Eq.~(\ref{eq:CRouth}), we have 
${\partial C_R}/{\partial p_\phi}= 0$ and ${\partial C_R}/{\partial p_\psi} = I_1/(I_3\rho)$,
and 
Eqs.~(\ref{eq:dCdp1})--(\ref{eq:dCdp2}) become
\begin{eqnarray*}
\xi^\phi - \left(\frac{f}{I_1}\right)(a\xi^\phi+\xi^\psi) &=& 0 \\
\xi^\psi + \left(\frac{fc}{I_1}+\frac{s^2}{I_3}\right)(a\xi^\phi+\xi^\psi) &=&
\left[\frac{I_1}{I_3}\right] \frac{1}{\rho} \,.
\end{eqnarray*}
Eliminating $\xi^\psi$ gives us an expression for $\xi^\phi$:
\begin{equation*}
\frac{1}{f} \left[ I_3 + s^2 + (I_3/I_1)f^2 \right] \xi^\phi = \frac{1}{\rho}  \,.
\end{equation*}
Simplifying this we get the infinitesimal Noether symmetry
\begin{equation}
\bfXi_R \equiv
\left(
\begin{matrix} \xi^\theta  \\  \xi^\phi  \\ \xi^\psi \\  \xi^X \\ \xi^Y \end{matrix}
\right)
=
\rho \left(
\begin{matrix}  0 \\  f  \\ (I_1-af)  \\  -I_1 s c_\phi  \\  - I_1 s s_\phi \end{matrix}
\right)\,.
\label{eq:Rgens}
\end{equation}
\section{Interpretation of the Routh sphere symmetries}
\label{sec:interp}

Each infinitesimal Noether symmetry associated with a constant of the motion has a
geometrical interpretation, obtained by integrating the it to construct a finite
transformation depending on one free parameter.  
Let us call this free parameter $\free$.

\subsection*{Jellett symmetry}

For the Jellett constant, the Noether symmetry (\ref{eq:Jgens}) leads to the equations
$$
\frac{\dd \theta}{\dd \free} = 0, \quad \frac{\dd \phi}{\dd \free} = 1, \quad \frac{\dd \psi}{\dd \free} = -a, \quad \frac{\dd X}{\dd \free} = 0, \quad \frac{\dd Y}{\dd \free} = 0\,.
$$
This has solution
$$
\theta=\theta_0, \quad X = X_0,  \quad Y = Y_0 \quad \text{(constants)},
$$
$$
\phi(\free) = \free + \phi_0, \quad \psi(\free) = - a\, \free + \psi_0\,.
$$
We consider the virtual motion corresponding to this free parameter $\free$.
The angular velocity is simply:
$$
\mbf{\Omega} = \frac{\dd \phi}{\dd \free} \tbf{K} + \frac{\dd \psi}{\dd \free} \mbf{k}
 =  \tbf{K} - a\,\mbf{k} = - \mbf{r} \,,
$$
where $\tbf{K}$ is the unit vector in the vertical direction in the inertial frame,
and $\mbf{k}$ is the unit vector in the body frame, pointing along the symmetry axis
of the body.
The \emph{contact vector} $\mbf{r}$ 
points from the centre of mass $\mbf{O}$ to the contact point $\mbf{P}$
(see Fig.~\ref{fig:RS}).  It follows that $\mbf{\Omega}$
is the vector pointing from the contact point $\mbf{P}$ to the centre of mass $\mbf{O}$.

Since the position of the centre of mass is fixed, while the Euler angle $\phi$ changes
at a constant rate, we deduce that the angular velocity $\mbf{\Omega}$ precesses
uniformly about the vertical axis $\tbf{K}$, describing a cone. The period of this
precession is $\Delta \free = 2\pi$, the same as the period of the angle $\phi$.
The period of the $\psi$ angle is $2 \pi a$, which is almost never commensurate with $2 \pi$. 
Hence, the motion is generically quasi-periodic.

\subsection*{Routh symmetry}

For the Routh constant, the Noether symmetry (\ref{eq:Rgens}) leads to the equations
\begin{equation}
\frac{\dd \theta}{\dd \tilde\free} = 0, \quad
\frac{\dd \phi}{\dd \tilde\free} = \rho f,
\quad \frac{\dd \psi}{\dd \tilde\free} = \rho (I_1 - a f),
\quad \frac{\dd X}{\dd \tilde\free} = -\rho I_1 s\,c_\phi,
\quad \frac{\dd Y}{\dd \tilde\free} = -\rho I_1 s\,s_\phi \,.
\label{eq:solro}
\end{equation}
Observing that, for $\theta$ constant, $\rho$ is a positive constant, we will use the
rescaled parameter $\rho \tilde\free$ as our free parameter $\free$ from here on.
We can solve the first three equations directly:
\begin{equation}
\theta = \theta_0  \quad \text{(constant)} \,,
\qquad \phi(\free) = f \free + \phi_0 \,,
\quad \psi(\free) = (I_1 - a f) \free + \psi_0 \,,
\label{eq:symmthphps}
\end{equation}
where $f$ depends on $\theta$ and is thus constant.
As in the case of the Jellett symmetry, the angles $\phi$ and $\psi$ change at constant rates,
with ratio $\dd\psi/\dd \phi = - a + I_1/f$, again incommensurate in general. As $\theta$ varies from $0$ to $\pi$,
this ratio may take arbitrary values outside the open interval $(-a - I_1/(1+a), -a + I_1/(1-a))$.
In particular, as $I_1 >0$ it follows $\dd\psi/\dd \phi \neq -a$ which shows that the Routh case
does not contain the Jellett case. 

Let us write the equations for $X$ and $Y$, the last two equations of (\ref{eq:solro}),
explicitly, using the partial solutions just found:
\begin{equation}
\frac{\dd X}{\dd \free} = -I_1 s\,\cos(f \free + \phi_0) \,,
\qquad \frac{\dd Y}{\dd \free} = -I_1 s\,\sin(f \free + \phi_0) \,.
\label{eq:symmXY}
\end{equation}
The solution to these is immediate: letting $(X_0, Y_0)$ be the value of $(X,Y)$
at $\free = 0$, we have
$$
X(\free) =  
-\frac{I_1 s}{f} \left[\sin(f \free + \phi_0) - \sin(\phi_0)\right]+ X_0\,,
\qquad Y(\free) = \frac{I_1 s}{f}\left[\cos(f \free + \phi_0) - \cos(\phi_0)\right] + Y_0\,.
$$ 
The interpretation of this solution is as follows:
\begin{itemize}
\item
If $f\neq0$ then the projection of the centre of mass onto the underlying plane
describes a circle of radius $R = {I_1 s}/{|f|}$, centred at
$\left(X_0 + ({I_1 s}/{f})\sin\phi_0, Y_0 - ({I_1 s}/{f})\cos\phi_0\right)$,
with period $\Delta \free = 2 \pi / |f|$.
Noting that $I_1$ and $s$ are non-negative, the sense of rotation of this circular motion
is positive if $f>0$ and negative if $f<0$. An interesting case is when the parameters
$I_1, a$ and the angle $\theta$ are such that $I_1 - a f = 0$, which requires $f>0$ in particular.
Then the ball does not spin with respect to its symmetry axis: $\psi(\free) = \psi_0$
for all $\free$, and thus the motion corresponds to the ball spinning in the positive sense
with respect to the vertical axis $\tbf{K}$: the vector $\mbf{k}$ along the body's symmetry
axis precesses about the vertical $\tbf{K}$ with period $\Delta \free$.
\item
If $f=0$, namely if we choose $\theta = \cos^{-1}a$ (which is always possible),
then there is no circular motion (the radius tends to infinity): the azimuthal angle $\phi$
is now constant while the ball spins and therefore the centre of mass moves on a straight line.
The solution of (\ref{eq:symmthphps}) and (\ref{eq:symmXY}) in this case is
$$
\phi = \phi_0,
\quad \psi(\free) = I_1 \free + \psi_0 \,
\quad X(\free)= - I_1 s \free \cos (\phi_0)  + X_0 \,
\quad Y(\free) = - I_1 s \free \sin (\phi_0)+ Y_0 \,.
$$ 
so the centre of mass moves in a straight line as the Routh sphere rolls.
\end{itemize}



\section{{Chaplygin ball on a rotating turntable}}
\label{sec:ChapBall}

The dynamics of a  Chaplygin ball on a rotating turntable were analysed in \cite{BBM18}. 
The centre of mass of the ball coincides with the geometric centre and $I_1 = I_2 \ne I_3$.
The holonomic constraint confines the geometric centre to remain at
unit distance above the underlying plane so that the vertical velocity
of the centre of mass vanishes. 

Assuming unit mass and unit radius for the Chaplygin ball, the Lagrangian is
\smallskip
\[
L = \half\bigl[
I_1\,\dot\theta^2 + (I_1 s^2+I_3 c^2 )\dot\phi^2 + (2I_3 c)\dot\phi\dot\psi +
(I_3)\dot\psi^2 + \dot X^2 + \dot Y^2 \bigr]  
\]
\smallskip\noindent
where $s=\sin\theta$, $c=\cos\theta$ as above.
The potential energy is constant and is taken to be zero.
We note that, as for the Routh sphere, $L$ is independent of 
both $\phi$ and $\psi$.

There are two nonholonomic constraints, which are linear and homogeneous in the velocities,
corresponding to rolling motion without slipping with respect to the rotating turntable:
\begin{eqnarray}
\dot X &=& \phantom{-} s_\phi \dot\theta -  s c_\phi \dot\psi - \Omega Y
\label{eq:XdotCB} \\
\dot Y &=&          -  c_\phi \dot\theta -  s s_\phi \dot\psi + \Omega X
\label{eq:YdotCB}
\end{eqnarray}
where $c_\phi=\cos\phi$ and $s_\phi=\sin\phi$, as above,
and $\Omega$ is the (constant) angular velocity of the rotating turntable.
We write these constraints in the form 
$$
\gamma^\kappa \equiv A^\kappa_\mu \dot q^\mu +  B^\kappa_\mu q^\mu = 0
$$
where $q^\mu = \left(\theta,\phi,\psi,X,Y\right)$ and
$\dot q^\mu = \left(\dot\theta,\dot\phi,\dot\psi,\dot X, \dot Y\right)$. Thus,
$$
A^\kappa_\mu = 
\left[
\begin{matrix}
-  s_\phi & 0 &  s c_\phi & 1 & 0 \\
   c_\phi & 0 &  s s_\phi & 0 & 1
\end{matrix}
\right]
\,\qquad\mbox{and}\qquad
B^\kappa_\mu = 
\left[
\begin{matrix}
           0    & 0 &     0   &    0     &  \Omega  \\
           0    & 0 &     0   & -\Omega  & 0
\end{matrix}
\right]
\,.
$$


\newcommand{\bLC}{\mathbf{L}_\mathrm{C}}
\newcommand{\bLP}{\mathbf{L}_\mathrm{P}}


We describe a {\bf systematic method} to find Noether symmetries and corresponding
constants for the Chaplygin ball on a rotating turntable.
\begin{enumerate}
\item We require the symmetries to satisfy the nonholonomic constraints 
(\ref{eq:XdotCB}) and (\ref{eq:YdotCB}):
\begin{eqnarray}
- s_\phi \xi^\theta + s c_\phi \xi^\psi + \xi^X + \Omega Y \tau &=& 0  \,,
\label{eq:con1} \\
  c_\phi \xi^\theta + s s_\phi \xi^\psi + \xi^Y - \Omega X \tau &=& 0 \,.
\label{eq:con2}
\end{eqnarray}
Note that the component $\xi^\phi$ is absent from these equations.
The constraints are two linear algebraic equations for the six symmetry components
$(\xi^\theta, \xi^\phi, \xi^\psi, \xi^X, \xi^Y, \tau)$, which reduce the number
of independent symmetry components to four.  
\item We make an \emph{ansatz} for some symmetry components. For example, we might
require $\xi^\phi$ to be the only non-vanishing component.
\item In the invariance identity (\ref{eq:InvId1}), we substitute the symmetry components
that are known from the ansatz. This yields a differential equation for the remaining
symmetry components.
\item We solve the equation for these components. We can then construct the corresponding
conserved quantities, using the invariance identity in the form (\ref{eq:InvId2}).
\end{enumerate}

\subsection*{Symmetry for the vertical component of angular momentum}

Noting that $\xi^\phi$ does not occur in the nonholonomic constraints
(\ref{eq:con1}) and (\ref{eq:con2}), we seek a symmetry
$$
(\xi^\theta, \xi^\phi, \xi^\psi, \xi^X, \xi^Y, \tau) = 
(    0     , \xi^\phi,      0  ,  0   ,   0  ,  0  )   \,.
$$
This symmetry automatically satisfies the nonholonomic constraints. 
Now, because $\phi$ is an ignorable coordinate, the invariance identity
(\ref{eq:InvId1}) becomes $p_\phi \dot\xi^\phi = 0$, with solution $\xi^\phi = $ constant.
Then the invariance identity in the form (\ref{eq:InvId2})
becomes $\dd p_\phi/\dd t = 0$ so the $\phi$-component of angular momentum 
\begin{equation}
L_Z \equiv p_\phi
\label{eq:IntLZ}
\end{equation}
is an integral of the motion.

\subsection*{Symmetries for horizontal components of angular momentum}

Noting the unit coefficients of $\xi^X$ annd $\xi^Y$ in
constraints (\ref{eq:con1}) and (\ref{eq:con2}), we seek two types of symmetry:
\begin{eqnarray}
(\xi^\theta, \xi^\phi, \xi^\psi, \xi^X, \xi^Y, \tau) &=& 
(\xi^\theta,     \xi^\phi   , \xi^\psi,   1  ,   0  ,   0 ) \,,
\label{eq:xiX} \\
(\xi^\theta, \xi^\phi, \xi^\psi, \xi^X, \xi^Y, \tau) &=& 
(\xi^\theta,     \xi^\phi   , \xi^\psi,   0  ,   1  ,   0 ) 
\label{eq:xiY} \,.
\end{eqnarray}
We consider these symmetries in turn. Substituting (\ref{eq:xiX}) in the constraints,
we easily solve for $\xi^\theta$ and $\xi^\psi$:
\begin{equation}
\xi^\theta  = s_\phi \,, \qquad \xi^\psi  = -c_\phi/s \,.
\label{eq:xithetaphi}
\end{equation}
These immediately give us expressions for 
$\dot\xi^\theta$ and $\dot\xi^\psi$:
$$
\dot\xi^\theta  = c_\phi \dot\phi \,, \qquad
\dot\xi^\psi  = (c c_\phi /s^2)\dot\theta + (s_\phi/s)\dot\phi  \,.
$$
Using these in the invariance identity (\ref{eq:InvId1}),
which is $(\partial L/\partial\theta)\xi^\theta + p_\mu \dot\xi^\mu = 0$,
we obtain an equation for $\dot\xi^\phi$:
$$
\dot\xi^\phi = -\left( \frac{c_\phi}{s^2} \dot\theta + \frac{c s_\phi}{s} \dot\phi \right)
\,.
$$
This implies that $\xi^\phi$ is a function of $\theta$ and $\phi$ only. We obtain
$$
\frac{\partial\xi^\phi}{\partial\theta} = -\frac{c_\phi}{s^2} \,, \qquad
\frac{\partial\xi^\phi}{\partial\phi} = -\frac{c s_\phi}{s} \,.
$$
These are easily seen to satisfy the compatibility condition
${\partial^2\xi^\phi / \partial\theta\partial\phi} =
 {\partial^2\xi^\phi / \partial\phi\partial\theta}$
and we immediately have the solution
\begin{equation}
\xi^\phi = \frac{c c_\phi}{s} \,.
\label{eq:xipsi}
\end{equation}
The final step is to substitute (\ref{eq:xithetaphi}) and (\ref{eq:xipsi}) into
the invariance identity (\ref{eq:InvId2}) to obtain the Noether integral
\begin{equation}
L_Y \equiv s_\phi p_\theta + \left(\frac{c c_\phi}{s}\right) p_\phi
                - \left(\frac{c_\phi}{s}\right) p_\psi + p_X \,.
\label{eq:int2}
\end{equation}
A similar analysis starting from symmetry (\ref{eq:xiY}) yields the Noether integral
\begin{equation}
L_X \equiv c_\phi p_\theta - \left(\frac{c s_\phi}{s}\right) p_\phi
                + \left(\frac{s_\phi}{s}\right) p_\psi - p_Y \,.
\label{eq:int3}
\end{equation}

\subsection*{Symmetry for an integral involving the energy}

To obtain integrals which are non-linear in the velocities, we need to 
assume $\tau \neq 0$. We seek a symmetry such that
$$
(\xi^\theta, \xi^\phi, \xi^\psi, \xi^X, \xi^Y, \tau) = 
(    0     ,     0   ,      0  , \xi^X, \xi^Y, \tau)   \,.
$$
The constraints (\ref{eq:con1}) and (\ref{eq:con2}) then become
\begin{eqnarray}
\xi^X + \Omega Y\, \tau &=& 0 \,,
\label{eq:xiconX} \\
\xi^Y - \Omega X\, \tau &=& 0 \,.
\label{eq:xiconY}
\end{eqnarray}
The invariance identity (\ref{eq:InvId1}) is then
$$
p_X \dot\xi^X + p_Y \dot\xi^Y - H \dot\tau = 0\,.
$$
Differentiating the constraints (\ref{eq:xiconX}) and (\ref{eq:xiconY}) and
substituting for $\dot\xi^X$ and $\dot\xi^Y$, we get
$$
[H - \Omega(X \dot Y - \dot X Y ) ] \dot\tau = 0\,.
$$
This is satisfied for constant $\tau$.
Therefore, the invariance identity (\ref{eq:InvId2}) gives us the integral
\begin{equation}
J \equiv H - \Omega L_{\mathrm{O}} \,,
\label{eq:Jint}
\end{equation}
where
$L_{\mathrm{O}} \equiv (X \dot Y - \dot X Y ) = \bK\bdot(\mathbf{R}\boldsymbol{\times}\mathbf{\dot R})$
is the the angular momentum due to the centre of mass about the origin of the space frame
($\mathbf{R}=X\mathbf{I}+Y\mathbf{J}$ is the position vector
of the point of contact in the space frame).


\subsection*{Physical interpretation of the integrals}

The angular momentum about the centre of mass is
$$
\bLC
\equiv \mathbb{I}_\mathrm{C} \boldsymbol{\omega}
= I_1\omega_1 + I_2\omega_2 + I_3\omega_3\,.
$$
Following \cite{BBM18}, we compute the angular momentum about 
the point of contact, which is, in our notation,
$$
\bLP =
\mathbb{I}_\mathrm{C} \boldsymbol{\omega}
+  \mathbf{K}\boldsymbol{\times}(\boldsymbol{\omega\times}\mathbf{K}) 
- \Omega\mathbf{R} \,.
$$
We note that both the second and third terms on the right are horizontal vectors.

It was shown by \cite{BBM18} that, in the body frame,
\begin{equation}
\left(\frac{\dd\bLP}{\dd t}\right)_\mathrm{B} = \bLP\boldsymbol{\times\omega}.
\label{eq:LP}
\end{equation}
Therefore, in the space frame, 
$$
\left(\frac{\dd\bLP}{\dd t}\right)_\mathrm{S} =
\left(\frac{\dd\bLP}{\dd t}\right)_\mathrm{B} + \boldsymbol{\omega\times}\bLP
=\boldsymbol{0}\,.
$$
It therefore follows that
$F_1=\mathbf{I}\boldsymbol{\cdot}\bLP$, 
$F_2=\mathbf{J}\boldsymbol{\cdot}\bLP$ and 
$F_3=\mathbf{K}\boldsymbol{\cdot}\bLP$ are integrals of the motion.
Computation of $F_3$ is simple, since
only the first term of (\ref{eq:LP}) contributes:
$\mathbf{K}\boldsymbol{\cdot}\bLP = p_\phi$.
Expressions for the remaining integrals can be computed:
\begin{eqnarray*}
\mathbf{I}\boldsymbol{\cdot}\bLP &=&
c_\phi p_\theta - \left(\frac{c s_\phi}{s}\right) p_\phi
+ \left(\frac{s_\phi}{s}\right) p_\psi -  p_Y \,,\\
\mathbf{J}\boldsymbol{\cdot}\bLP &=&
s_\phi p_\theta + \left(\frac{c c_\phi}{s}\right) p_\phi
- \left(\frac{c_\phi}{s}\right) p_\psi +  p_X \,.
\end{eqnarray*}
We see  that the three components of
$\bLP$ in the space frame are precisely the three integrals
$(L_X, L_Y,L_Z)$ that we have derived from Noether's theorem.


In \cite{BBM18}, another integral, similar to the Jacobi integral, was found:
$$
E = \half \bom \bdot \bigl[ \mathbb{I}_\mathrm{C}\bom + \bK\bX(\bom\bX\bK) \bigr]
       - \half\Omega^2(X^2+Y^2) \,.
$$
They cite the origin of this integral as \cite{FGS18}. It is straightforward to show that
$E$ is identical to the integral $J$ in (\ref{eq:Jint}), which we found using Noether's
theorem.

\subsection*{Interpretation of the symmetries}
We proceed as in Section \ref{sec:interp} to find finite versions of the four
infinitesimal symmetries just found. 

\noindent $\bullet$ \textbf{Symmetry for $L_Z$.}
In terms of the free parameter $\alpha$ of the symmetry, we get the equation
$$\frac{d\phi}{d\alpha} = 1\,,$$ 
with solution $\phi(\alpha) = \alpha$. The remaining coordinates $(\theta, \psi, X, Y)$ are kept constant.
This corresponds geometrically to the spinning of the ball about the point of contact,
at a constant angular velocity ${d\phi}/{d\alpha} = 1$.  

\noindent $\bullet$ \textbf{Symmetry for $L_Y$.}
Reading off the coefficients of $p_\mu$ from equation (\ref{eq:int2}), we get the equations
$$
\frac{d \theta}{d\alpha} = \sin \phi\,,
\qquad \frac{d \phi}{d\alpha} = \frac{\cos \theta \cos \phi}{\sin \theta} \,,
\qquad \frac{d \psi}{d\alpha} = - \frac{\cos \phi}{\sin \theta} \,,
\qquad \frac{d X}{d\alpha} = 1, \qquad \frac{d Y}{d\alpha} = 0 \,.
$$
One immediately gets $X(\alpha) = \alpha$ and $Y=\text{constant}$.
This suggest that the geometric interpretation of this symmetry corresponds to a
rotation of the ball such that $Y$ is constant and $X$ changes linearly. 
To see this, consider the equations for $\theta$ and $\phi$. They provide the first integral
$$
\sin \theta \cos \phi = y_0\quad \text{(constant)},
$$
which validates this interpretation.  A less obvious result follows from the equation
for $\psi$, which can be solved by quadrature, giving the implicit first integral
$$
\tan(\psi-\psi_0) = \cos\theta \cot \phi \qquad \qquad (\psi_0 =  \text{constant}) \,.
$$ 

\noindent $\bullet$ \textbf{Symmetry for $L_X$.}
Reading off the coefficients of $p_\mu$ from equation (\ref{eq:int3}), we get the equations
$$
\frac{d \theta}{d\alpha} = \cos \phi \,,
\qquad \frac{d \phi}{d\alpha} = -\frac{\cos \theta \sin \phi}{\sin \theta} \,,
\qquad \frac{d \psi}{d\alpha} =  \frac{\sin \phi}{\sin \theta} \,,
\qquad \frac{d X}{d\alpha} = 0 \,,
\qquad \frac{d Y}{d\alpha} = -1 \,.
$$
Here the interpretation of the symmetry corresponds to a rotation of the ball
such that $X$ is constant and $Y$ changes linearly. In a similar fashion to the results
obtained for $L_Y$, we obtain the following first integrals:
$$
\sin \theta \sin \phi = x_0 \quad \text{(constant)},
\qquad\quad \tan(\psi-\psi_0) = - \cos\theta \tan \phi
\qquad\quad (\psi_0 =  \text{constant}) \,.
$$


\section{Discussion}

The key property of the infinitesimal Noether symmetries found for the
Routh sphere and the Chaplygin ball is that they respect the nonholonomic constraints.
In the more general case of the {Chaplygin top} or the Rock'n'roller,
it is not known whether an infinitesimal Noether symmetry that respects
the nonholonomic constraints exists. If such a symmetry existed, then a
constant of motion could be constructed via equation (\ref{eq:Noether1}).

For example, it is possible to show for these more general cases that the transformation
$$\phi \to \phi + \epsilon$$ (while keeping all other variables unchanged, including $X$ and $Y$)
is an infinitesimal Noether symmetry. However, this symmetry does not respect the
nonholonomic constraints (\ref{eq:Xdot})--(\ref{eq:Ydot}) (with velocities replaced
by the generators). In fact, from equation (\ref{eq:InvId2}) we obtain
$$\frac{\dd p_\phi}{\dd t} = a s (\lambda_1 c_\phi + \lambda_2 s_\phi)\,,$$
where $\lambda_1$ and $\lambda_2$ are the multipliers associated with the constraints
(\ref{eq:Xdot}) and (\ref{eq:Ydot}) respectively. This example shows that a Noether symmetry
is potentially useful even if it does not respect the nonholonomic constraints: it provides
direct formulas for the total time derivative of quantities, which in principle could be
exploited for applications such as finding Lyapunov functions. 

{Another avenue of research is the understanding of the Lie algebra
between the Noether symmetries that we found for nonholonomic systems. In the case of
holonomic systems, it is well known that the Lie bracket between two symmetries is another symmetry.
This leads to a method for finding new integrals starting from known ones \cite{BH03}.
However, when nonholonomic constraints are imposed, the usual Lie bracket between two
Noether symmetries does not necessarily produce another Noether symmetry.  Further research
on the relation between Poisson brackets and symmetries (see \cite{Cushman98, BandT13}
for studies in the context of the Routh sphere), is needed to generalise the Lie bracket
as a method to produce new Noether symmetries.}


\section*{Acknowledgement}

We thank the reviewers for valuable comments, which have helped us to improve the paper.
We are grateful to Vakhtang Putkaradze for fruitful discussions about
Noether's Theorem and its use in the analysis of integrable systems. 





\end{document}